\documentclass[12pt]{JHEP3}
\usepackage{amsmath,bbm}

\title{Boundary operators in minimal Liouville gravity and matrix models}
\author{Jean-Emile Bourgine and Goro Ishiki\\ Center for Quantum Spacetime (CQUeST)\\ Sogang University, Seoul 121-742, Korea\\ E-mail: \email{jebourgine@sogang.ac.kr} and \email{ishiki@post.kek.jp}}
\author{Chaiho Rim\\ Department of Physics and Center for Quantum Spacetime (CQUeST)\\ Sogang University, Seoul 121-742, Korea\\ E-mail: \email{rimpine@sogang.ac.kr}}
\abstract{
We interpret the matrix boundaries of the one matrix model (1MM) recently constructed by two of the authors as an outcome of a relation among FZZT branes. In the double scaling limit, the 1MM is described by the $(2,2p+1)$ minimal Liouville gravity. These matrix operators are shown to create a boundary with matter boundary conditions given by the Cardy states. We also demonstrate a recursion relation among the matrix disc correlator with two different boundaries. This construction is then extended to the two matrix model and the disc correlator with two boundaries is compared with the Liouville boundary two point functions. In addition, the realization within the matrix model of several symmetries among FZZT branes is discussed.}

\preprint{}

\let\refOld\ref
\renewcommand{\ref}[1]{(\refOld{#1})}

\newcommand{\tr}{\text{tr }}

\newcommand{\bL}{b}

\newif\ifdefinitif

 \def\d{\delta}

 \def\p{\partial}

 \def\b{\beta}
 
 \def\d{\delta}
 \def\e{\epsilon}

 \def\th{\theta}
 
 \def\k{\kappa}

 \def\x{\xi}
 \def\r{\rho}
 
 \def\s{\sigma}
 
 \def\th{\theta}
 \def\z{\zeta }

 \def\o{\omega }
 
\def\equskip{\!\!\!\!\!\!\!\!} 
\def\la{\left\langle}
\def\ra{\right\rangle}

\def\Op{\mathcal{O}}
\def\tOp{\tilde{\mathcal{O}}}

\def\tW{\tilde{W}}
\def\tw{\tilde{\o}}
\def\ts{\tilde{s}}

\definitiftrue
\begin{document}
\section{Introduction}
Since their introduction in the middle 80's in the context of string theory \cite{Kawai1985,Onogi1988,Ishibashi1988} and lattice statistical models, \cite{Cardy1984,Cardy1986,Saleur1989}, the role played by the boundary conditions (BC) in 2D conformal field theories (CFT) has been widely uncovered \cite{Cardy1989,Cardy1991}. However, several intriguing questions remain, the most challenging involving the interplay between boundaries and renormalization group (RG) flows. This includes the description of RG flows driven by perturbations of conformal boundary conditions \cite{Dorey2009}, and the matching of boundary conditions between two CFTs related under a bulk RG flow \cite{Fredenhagen2009}. Both attempts to tackle these problems, which rely respectively on integrability and perturbative RG analysis, make use of the $g$-function to identify the boundary conditions. Unfortunately, the results obtained are restricted to the thermal flows of the unitary $(p,p+1)$ minimal models, more precisely to the tricritical to critical Ising flow in \cite{Dorey2009}, and to a pertubative expansion for large $p$ in \cite{Fredenhagen2009}. Very recently, these results have been extended to any $p$ under plausible assumptions, using the staircase model \cite{Dorey2010}.

We focus here on a very promising alternative approach based on matrix models. Matrix models can be used to construct statistical models defined on a dynamical lattice and allow to take an explicit continuum limit. In this double scaling limit, the critical points are described by a CFT, referred as ``matter'', coupled to a two dimensional quantum metric. In the conformal gauge, the metric field reduces to a scalar field that obeys the Liouville action, and the theory is called ``Liouville gravity''. The matter CFT is nothing but the theory that describes the scaling limit of the statistical model on a fixed lattice at its critical point. The study of the Liouville gravity correlation functions allows to recover several features of the bare matter CFT by making use of the KPZ relations \cite{Knizhnik1988,David1988,Distler1988} between the central charges and operator dimensions of matter and Liouville theories. Since the matrix model description is also valid away from the critical point, it reveals a profitable way to understand both bulk and boundary RG flows. This approach has been successfully applied to the lattice $O(n)$ model whose continuum limit consists in a set of non-rational CFT which interpolates between the unitary minimal series. There, the matrix model gave a description of the thermal bulk \cite{Kostov2006} and boundary \cite{Bourgine2010} flows, as well as the boundary anisotropic flow \cite{Bourgine2010,Bourgine2009,Bourgine2009a}. The goal of the present article is to pave the way for a similar construction within the two matrix model (2MM). Taking a suitable double scaling limit, this model is known to describe a $(p,q)$ minimal model coupled to the Liouville field, sometimes referred as $(p,q)$ minimal Liouville gravity.

In the Liouville gravity framework, the boundary condition changing operators (or bcc operators) $\psi_a$ of the matter theory are dressed by the Liouville field $\varphi$: the operators are multiplied by a boundary vertex operator of the Liouville theory, $B_a=\psi_a e^{\b_a \varphi}$. In order to get a diffeomorphism invariant operator, the coordinates should also be integrated over the boundary of the worldsheet. In the following, we will consider the matrix model in the planar limit, thus restricting to worldsheets with the topology of a disc. The Liouville charge $\b_a$ is related to the dimension $h_a$ of the matter boundary operator through the KPZ formula. In the case of the $(p,q)$ minimal models ($p<q$ for definiteness), the bcc operators fill the Kac table: $a=(k,l)$ with $1\leq k<p$ and $1\leq l<q$. In addition, $\psi_{k,l}$ and $\psi_{p-k,q-l}$ should be identified since they wear the same scaling dimensions
\begin{equation}
h_{k,l}=h_{p-k,q-l}=\dfrac14\left[(k/\bL-l\bL)^2-(1/\bL-\bL)^2\right],\qquad \bL^2=p/q\ .
\end{equation}
The KPZ equation is quadratic in $\b_a$, leading to two different solutions, or ``dressings'',
\begin{equation}\label{L_charge}
\b_{k,l}^\pm=\dfrac{Q}{2}\mp|P_{k,l}|\ ,\qquad P_{k,l}=\dfrac{k}{2\bL}-\dfrac{\bL l}{2}\ ,\qquad Q=\bL+\dfrac1\bL\ .
\end{equation}
Both dressings can be realized in the 2MM and we will see in particular that the symmetry $(k,l)\to(p-k,q-l)$ involve an exchange of the dressings for the boundary two point functions. However, when not mentioned explicitly, the dressing will be supposed to satisfy the Seiberg's bound \cite{Seiberg1990}, $\b=\b^+$. We will come back to this subtle point below.

The boundary states must specify the boundary conditions for the matter and the Liouville fields. Called FZZT branes \cite{Fateev2000,Teschner2000}, they are constructed as a tensor product of a Cardy brane $|k,l>$ with a Liouville boundary state $|s>$. The boundary parameter $s$ is related to the cosmological constant $\xi(s)$ which control the length of the boundary. The presence of conformal Killing vectors allow to fix the position of three operators on the boundary. As a consequence of the absence of interaction terms between the Liouville and matter fields in the action, both contributions to the boundary one, two and three point functions factorize. The Liouville correlator is the only part that contains a dependency in the cosmological constants. Most of the results of this paper exploits this property of the boundary two point functions. More precisely, the Liouville momentum $P_{k,l}$ can be identified by comparing the cosmological constant dependence of the matrix model correlators in the continuum limit and the expression for the Liouville boundary two point function which has been derived in \cite{Fateev2000}. This identification provide a way to determine the dimension of the bcc operator inserted between two given boundaries.

The first step of this ambitious program to study boundary RG flows is to construct the equivalent of the Liouville gravity bcc operators within the matrix model. This has been recently achieved in the one matrix model (1MM) \cite{Ishiki2010} and will be reviewed in the first section. This matrix model can be tuned to give a description of $(2,2p+1)$ minimal Liouville gravity in the double scaling limit. The aim of the present article is to discuss further this construction and to extend it to the 2MM case (section 2) in order to recover the bcc operators of general $(p,q)$ minimal Liouville gravity. The strategy we employed is first to introduce the matrix operator that creates a boundary inside a single boundary correlator, motivated by a decomposition relation among FZZT branes \ref{rel_FZZT_2MM}. Then, we turn to two boundary matrix correlator and determine the bcc operators inserted from the identification with the boundary Liouville two point function, as explained above. We eventually deduce the matter boundary conditions from a CFT argument. For a matter of clarity, most of the calculations are confined to the appendix. It contains in particular the demonstration of the two main results: the recursion relation \ref{recursion} for the 1MM disc correlators with two general boundaries already mentioned in \cite{Ishiki2010} but not proven, and the identification of the 2MM correlation functions \ref{def_mixed} with the corresponding Liouville boundary two point function. Additional remarks and perspectives are gathered in the conclusion.

\section{One matrix model}
The 1MM partition function is defined as an integral over the hermitian matrix $M$ of size $N\times N$, with an action given by a polynomial potential $V(M)$\footnote{In the following we consider \ref{def_1MM} as a formal matrix model, i.e. the matrix integrals are seen as formal series which terms are computed using the gaussian kernel.},
\begin{equation}\label{def_1MM}
Z=\int{D[M]\ e^{-\frac{N}{g}\tr{V(M)}}}.
\end{equation}
The free energy $N^{-2}\log{(Z)}$ is a generating function of discretized closed surfaces where the weight of the polygons is given by the polynomial coefficients of the corresponding degree. In the large $N$ expansion of $\log{(Z)}$, each surface carries an overall factor $\k^{-A}N^\chi$ depending on its topology through the Euler characteristic $\chi$, and on its area $A$ with a cosmological constant $\k^2=1/g$. The planar limit consists in sending $N$ toward infinity, keeping only the surfaces with the sphere topology. The main quantity of the model is the spectral density associated to the matrix $M$, it is usually given through its Stieltjes transform, the resolvent $W(x)$:
\begin{equation}\label{resol}
W(x)=\dfrac{g}{N}\la\tr\dfrac1{x-M}\ra=\dfrac{V'(x)}{2}+\o(x)\ .
\end{equation}
In the planar limit, the resolvent describes a sum over discretized surfaces with the disc topology and one point marked on the boundary. The boundary cosmological constant $x$ controls the length of the boundary, as can be seen from the $x\to\infty$ expansion. When the number $N$ of eigenvalues goes toward infinity, the support of the density condensate into a continuous subset of the real line and the resolvent develops a branch cut. For our purpose, we concentrate on the single cut case where the support is an interval (for a review, see \cite{DiFrancesco1995,Ginsparg1993}).

The continuum limit is achieved by sending the bulk and boundary cosmological constants to their critical values $\k^*$ and $x^*$. Note that by shifting the matrix, the potential can always be chosen such that the critical value of $x$ is zero. The minimal $(2,2p+1)$ Liouville gravity is obtained after tuning the coefficients of the potential up to the degree $p$ in the double scaling limit. At the critical point, the mean area and boundary length diverge and we define the appropriate renormalized cosmological constants $\e^{2}\mu=\k-\k^*$ and $\e\xi=x$, with $\bL^2=2/(2p+1)$. The singular part of the resolvent\footnote{The non critical (or non universal) part, which is always polynomial, can be related to degenerate configurations and shall be discarded in the continuum limit, see \cite{Alexandrov2005}.} $\e^{1/\bL^2}\o(\xi)=\o(x)$ defined in \ref{resol} is identified to the one point function of the boundary operator $B_{1,1}$ on the disc. The renormalized quantity $\xi$ corresponds to the Liouville boundary cosmological constant, sometimes denoted $\mu_B$ in the literature \cite{Fateev2000}. In order to resolve the branch cut of the resolvent on $]-\infty,-u]$, $\xi$ is usually parameterized by the boundary parameter $s$,
\begin{equation}
\xi(s)=u\cosh{(\pi\bL s)}\ ,\qquad \o(\xi)=u^{1/\bL^2}\cosh{(\pi s/\bL)}\ .
\end{equation}
where $u=\sqrt{\mu/\sin(\pi\bL^2)}$. We will assume that the matter carries $(1,1)$ BC on the disc, which will be confirmed by the consistency of the construction below.

\subsection{Construction of the matrix boundaries}
In \cite{Ishiki2010}, non-trivial matrix boundaries were introduced using an additional vector field which couple to the matrix $M$. We use here another approach based on a relation among FZZT branes first noticed in \cite{Seiberg2004}, and derived in \cite{Basu05} from the boundary ground ring technique. This relation decomposes the FZZT brane $|s;1,l>$ as a sum over FZZT branes $|s_n;1,1>$ with trivial matter BC and shifted boundary parameter $s_n=s+in\bL$,
\begin{equation}\label{rel_FZZT}
|s;1,l>=\sum_{n=-(l-1),2}^{l-1}{|s_n;1,1>}
\end{equation}
where the index $n$ runs from $-(l-1)$ to $(l-1)$ with an increment of two. The relation \ref{rel_FZZT} trivially extends to the disc partition functions of the Liouville gravity $Z_\text{LG}(s;1,l)$. The resolvent is the derivative of the disc partition function $Z_\text{LG}(s;1,1)$ with respect to the cosmological constant. Consequently, the partition function $Z_\text{LG}(s;1,1)$ is obtained by integration of \ref{resol}, i.e. by taking the double scaling limit of $\frac{g}{N}\la\tr\log{(x-M)}\ra$. Applying the linear relation \ref{rel_FZZT} to the logarithmic correlator leads to regard the following quantity as a candidate for the more general $Z_\text{LG}(s;1,l)$ disc partition functions,
\begin{equation}
Z_l(\{x\}_l)=\sum_{n=-(l-1),2}^{l-1}{\dfrac{g}{N}\la\tr\log{(x_n-M)}\ra}\ ,\qquad x_n=\e u\cosh{(\pi\bL s_n)}\ .
\end{equation}
In this way, we recover the partition function considered in \cite{Ishiki2010} where was defined the matrix operator $F_l(M)$ as
\begin{equation}
Z_l(\{x\}_l)=\dfrac{g}{N}\la\tr\log{F_l(M)}\ra,\qquad F_l(M)=\prod_{n=-(l-1),2}^{l-1}{(x_n-M)}\ .
\end{equation}
We will show that the operator $F_l$ creates a boundary with cosmological constant $x=\e u\cosh{(\pi\bL s)}$ and matter boundary conditions $(1,l)$. First, we notice that the derivative with respect to $x$ of the disc partition function $Z_l(\{x\}_l)$ is proportional to the resolvent, and thus corresponds to the $B_{1,1}$ boundary one point function,
\begin{equation}\label{res_1pt}
\Op_l(\{x\}_l)=\dfrac{g}{N}\p_x\la\tr\log{F_l(M)}\ra=(-1)^{l-1}[l]_+\o(x)+(\text{n.u.})\ ,\qquad [l]_+=\dfrac{\sin{(\pi \bL^2 l)}}{\sin{(\pi \bL^2)}}\ .
\end{equation}
Here $(\text{n.u.})$ refers to the non universal polynomial part of the correlator. This confirms the identification of the cosmological constant $x$. The assertion concerning the $(1,l)$ matter boundary conditions is supported by taking the ratio\footnote{The ratio allows to neglect the contributions of leg factors and coordinates. Moreover, as the Liouville part does not depend on the matter BC but only on the dimension of the matter boundary operators, it also simplifies from this ratio.} of the singular part of $\Op_l$ and of the resolvent \cite{Ishiki2010}. We thus obtain the factor $(-1)^{l-1}[l]_+$ which is exactly the ratio of the matter parts $\la \psi_{1,1}\ra_{(1,l)}/\la \psi_{1,1}\ra_{(1,1)}$. Hence, the correlator $\Op_l$ corresponds in the double scaling limit to the $B_{1,1}$ boundary one point function with $\xi(s)$ cosmological constant and $(1,l)$ matter BC. This elucidates the role played by $F_l$ at the level of the disc with a single boundary.

In obtaining \ref{res_1pt}, it is crucial that the resolvent take the same value over the shifted cosmological constants $x_n$, $\o(x_n)=(-1)^{l-1}\o(x)$. This condition restricts to at most $p$ different solutions for $x_n$ so that $l\leq p$. Note that, as supposed by the identification of the boundary conditions $(1,l)$ and $(1,2p+1-l)$, the result \ref{res_1pt} is invariant under the exchange $l\to 2p+1-l$.\\

\subsection{Matrix correlators and Liouville boundary two point functions}
We investigate here the role played by the matrix operator $F_l$ at the level of a disc with two boundaries. More precisely, we present a method to determine the matter boundary conditions associated to the boundary created by $F_l$, based on the study of the Liouville boundary two point function. This analysis will be extended to the 2MM case in the next section. The idea is to consider the following correlator, interpreted as a disc with two boundaries constructed with respectively $F_1$ and $F_l$,
\begin{equation}\label{Ol}
\Op_l(x,\{y\}_l)=\dfrac{g}{N}\la\tr\dfrac1{x-M}\dfrac1{F_l(M)}\ra\ .
\end{equation}
In the continuum limit, the singular part of $\Op_l$ is proportional to the Liouville boundary two point function of $B_{1,l}$ operators, as proven in \cite{Ishiki2010}. The matrix operator $F_1$ creates a boundary with BC $(1,1)$ for the matter field. The matter bcc operator must belong to the fusion algebra of the Verma module associated with the two nearby boundary conditions. Any BC fused with the identity reproduces itself, thus, on the boundary created by $F_l$, the matter field BC wear the same label as the boundary operator: $(1,l)$. This identification is compatible with the symmetry $l\to2p+1-l$, provided we exchange the dressings by the Liouville field. We will come back to this point in the next section, dealing with a slightly more general case within the 2MM.

Once the role of the $F_l$ matrix operator has been elucidated, it is possible to understand the general two boundary matrix correlators
\begin{equation}
\Op_{l,k}(\{x\}_l,\{y\}_k)=\dfrac{g}{N}\la\tr\dfrac{1}{F_l(M)}\dfrac1{F_k(M)}\ra\ .
\end{equation}
The correlator $\Op_{l,k}$ corresponds to a disc with $(1,l)$ and $(1,k)$ matter BC in the continuum limit. These matter BC allows several bcc operator insertion, namely $\psi_{1,n}$ where $n=1+|l-k|\cdots n_\text{max},2$ and $n_\text{max}=\min\{1+l+k,4p+1-l-k\}$. In order to determine the bcc operators involved in $\Op_{l,k}$, we employ the following recursion relation,
\begin{equation}\label{recursion}
[k]_+\Op_{l,k+1}(\{x\}_l,\{y\}_{k+1})=[l]_+\Op_{l+1,k}(\{x\}_{l+1},\{y\}_k)+(\text{n.u.})\ .
\end{equation}
This relation was already mentioned in \cite{Ishiki2010}, but not proven. Its demonstration is one of the main results of this paper, it can be found in the appendix \ref{app_D}. Using the relation \ref{recursion} to decrease the index $l$ up to one, we eventually come back to the known correlator $\Op_{l+k-1}$ defined in \ref{Ol}. The identification with the Liouville boundary two point function leads to determine the bcc operator $B_{1,l+k-1}$. We notice that this operator depends only on the sum $l+k$, implying that many different boundary conditions lead to the same bcc operator. This is the meaning of the recursion \ref{recursion}: both Liouville part of the gravitational correlators are the same whereas the factor $[l]_+/[k]_+$ reflects the different matter parts. Unfortunately, it is not known yet how to construct the matrix model correlator corresponding to the insertion of the other bcc operators between $(1,l)$ and $(1,k)$ matter BC. This is one of the major open questions we hope to answer in a near future.

Up to now, we have demonstrated that the matrix operator $F_l$ can be used to replace $F_1$ in the case of a disc with one or two boundaries, thus promoting $(1,1)$ to $(1,l)$ matter BC. We believe that this can be generalized to any correlator of any genus, with an arbitrary number of boundaries, possibly disconnected. Moreover, the relation \ref{rel_sum_prod} employed in the proofs above allow to decompose these $F_l$ boundaries as a sum over $F_1$ boundaries with shifted cosmological constants, similarly to \ref{rel_FZZT}. All these correlators involving only $F_1$ can be evaluated recursively using the method developed in \cite{Eynard2004a}. The main complexity in the computation of the general correlator lies in the evaluation of the sum over shifted boundary parameters.

\section{Two matrix model}
The two matrix model was first introduced in \cite{Kazakov1986} in order to describe the 2D Ising model on a fluctuating lattice. It was subsequently generalized to arbitrary polynomial potential. The partition function is defined as an integral over two $N\times N$ hermitian matrices $X$ and $Y$ with potentials $V(X)$ and $U(Y)$,
\begin{equation}
Z_\text{2MM}=\int{dXdY\ \exp{\left[-\dfrac{N}{g}\tr{\Big (}V(X)+U(Y)-XY{\Big )}\right]}}.
\end{equation}
Each matrix is associated to a site wearing a different color and the $XY$ propagator is orthogonal to an edge of the interface between the two colors. The potentials generate propagators and vertices relating sites of the same color, similarly to the 1MM. In the case of a quadratic potential $U(Y)$, the matrix $Y$ can be integrated out and we recover the 1MM. The main quantities are the two eigenvalue densities associated to the matrices $X$ and $Y$, so that we need to introduce the resolvents\footnote{For later convenience we included a minus sign in the definition of the resolvents.},
\begin{equation}
W(x)=-\dfrac{g}{N}\la\tr\dfrac1{x-X}\ra\ ,\qquad \tW(y)=-\dfrac{g}{N}\la\tr\dfrac1{y-Y}\ra\ .
\end{equation}
These matrix correlators describe a disc with a boundary of a single color, $X$ or $Y$, in the presence of one marked point.%
\footnote{See \cite{Itoyama} for the amplitudes describing cylinder and 
higher disconnected loop-boundaries with ordinary boundary conditions.
}
Again, we restrict ourselves to the planar limit and the resolvents are multivalued functions with a branch cut on the support of the densities. We denote their respective singular part $\o(x)=V'(x)+W(x)$ and $\tw(y)=U'(y)+\tW(y)$, and the bare (bulk) cosmological constant is $\k^2=1/g$.

In the double scaling limit, the coefficients of the potentials $V$ and $U$ are tuned up to degree $q$ and $p$ respectively, in order to achieve the $(p,q)$ critical behavior ($p<q$ for definiteness) \cite{Daul1993}. Using a shift of the matrices, we set the critical values of the boundary cosmological constant to be zero. We then define the renormalized cosmological constants $\e^{2p}\mu=\k-\k^*$, $\e^p\xi=x$ and  $y=\e^q\z$. The dependence in the cut-off as been slightly modified from the 1MM in order to emphasize the symmetry between $p$ and $q$ that exchanges the two matrices $X$ and $Y$. The boundary cosmological constants are conveniently parameterized as\footnote{Again, we slightly modify here the previous parameterization in order to emphasize the symmetry. The boundary parameters of 1MM and 2MM are thus related by $\bL s_\text{1MM}=s_\text{2MM}/q$, $s_\text{1MM}/\bL=s_\text{2MM}/p$. The variable $\z(t)$ is identifed with a dual cosmological constant, thus the role played by $\bL$ and $1/\bL$ are exchanged, $t_\text{1MM}/\bL=t_\text{2MM}/p$, $\bL t_\text{1MM}=t_\text{2MM}/q$.}
\begin{equation}\label{param_xy}
\xi(s)=2u^p\cosh{(\pi s/q)}\ ,\qquad \z(t)=2 u^q\cosh{(\pi t/p)}\ ,\qquad \bL^2=p/q\ .
\end{equation}
The resolvent $\o(x)$ is identified in the continuum limit to the $B_{1,1}$ boundary one point function of the minimal Liouville gravity, with $\xi(s)$ boundary cosmological constant and $(1,1)$ matter BC, in agreement with the results of the first section. The $Y$ matrix resolvent also corresponds to the minimal Liouville gravity $B_{1,1}$ one point function, but the boundary wears the dual cosmological constant $\z(s)$, sometimes denoted $\tilde{\mu}_B$ in the literature. The identification of the matter BC corresponding to $\tw(y)$ will be made below. Both resolvents can be easily expressed using the boundary parameters $s$ and $t$,
\begin{equation}
\o(\xi)=2 u^q\cosh{(\pi s/p)}\ ,\qquad \tw(\z)=2u^p\cosh{(\pi t/q)}\ .
\end{equation}

Most of the information concerning the 2MM can be encoded into a single object, the spectral curve defined by
\begin{equation}
E(x,y)=(x-U'(y))(V'(x)-y)-P(x,y)+g\ ,
\end{equation}
with the polynomial
\begin{equation}
P(x,y)=\dfrac{g}{N}\la\tr\dfrac{V'(x)-V'(X)}{x-X}\dfrac{U'(y)-U'(Y)}{y-Y}\ra.
\end{equation}
Since $E(x,y)$ is a polynomial, the equation $E(x,y)=0$ defines an algebraic curve in $\mathbb{C}\times \mathbb{C}$. As explained in the appendix \ref{app_B}, this curve is equivalent to the condition $x=\o(y)$, or by symmetry $y=\o(x)$, that characterizes the distribution of eigenvalues. In the continuum limit, using the parameterization \ref{param_xy}, its expression is remarkably simple (see appendix \ref{app_B} for an explicit derivation),
\begin{equation}\label{spec_cont}
E(x,y)=\e^{pq}u^{pq}\left[\cosh{(\pi s)}-\cosh{(\pi t)}\right].
\end{equation}
The spectral curve allows to determine recursively all the correlators of the model \cite{Eynard2005,Eynard2005a}. More specifically, it will be related here to the expression of the mixed traces studied below.

\subsection{Study of the mixed trace}
The boundary operators of the 2MM were first considered in connection with the FZZT branes of Liouville gravity in \cite{Hosomichi2008}. In this paper, the author focus on the mixed trace correlation function, given by
\begin{equation}\label{mixed_trace}
\mathcal{M}(x,y)=\dfrac{g}{N}\la\tr\dfrac1{x-X}\dfrac1{y-Y}\ra=1-\dfrac{E(x,y)}{(x-\tw(y))(y-\o(x))}\ .
\end{equation}
The derivation of this expression is reviewed in appendix \ref{app_MT}. In the combinatorial interpretation of the 2MM, the mixed trace correlator corresponds to a disc with two boundaries of different color, each boundary is constructed using a different matrix (see e.g. \cite{Eynard2005}). The singular part of $\mathcal{M}$ is defined in the double scaling limit as $\e^{pq-p-q}\mathcal{M}(\xi,\z)=\mathcal{M}(x,y)-1$ which writes, using the parameterization \ref{param_xy},
\begin{equation}\label{expr_m_cont}
\mathcal{M}(\xi(s),\z(t))=\dfrac{u^{pq-p-q}}{4}\dfrac{\cosh{(\pi s)}-\cosh{(\pi t)}}{\left[\cosh{(\pi s/q)}-\cosh{(\pi t/q)}\right]\left[\cosh{(\pi s/p)}-\cosh{(\pi t/p)}\right]}\ .
\end{equation}
In \cite{Hosomichi2008}, this expression was shown to coincide with the Liouville part of the gravitational $B_{p-1,1}$ boundary two point function with boundary parameters $s$ and $t-ipq$. From the supposition that the boundary created by $X$ matrices has matter BC $(1,1)$, we deduce the matter BC $(p-1,1)$ of the $Y$ boundary. This claim will be checked at the level of the one point functions in the section \refOld{sec_sym_FZZT} below.

\subsection{Analysis of the 2MM boundaries}
Since the 2MM can be seen as a generalization of the 1MM, the later corresponding to the case $p=2$, the analysis of the first section still remains valid. In particular, the matrix operator $F_l(X)$ leads to a boundary with $(1,l)$ matter BC and $\xi(s)$ cosmological constant. Note that these matter boundary conditions form a closed subset under fusion,
\begin{equation}
(1,l)\times (1,l')=\sum_{n=1-|l-l'|}^{n_\text{max}}{(1,n)},\quad n_\text{max}=\min(1+l+l',2q-l-l'-1).
\end{equation}
The decomposition of FZZT-branes \ref{rel_FZZT} found in \cite{Seiberg2004} can be generalized to the $(p,q)$ minimal Liouville gravity,
\begin{equation}\label{rel_FZZT_2MM}
|s;k,l>=\sum_{m=-(l-1),2}^{l-1}{\sum_{n=-(k-1),2}^{k-1}{|s_{mn};1,1>}}\ ,\qquad s_{mn}=s+imp+inq\ .
\end{equation}
As in the one matrix model, the operator $F_l(X)$ provides the case $k=1$. The dual case $l=1$ is introduced as a new matrix operator,
\begin{equation}
F_k(Y)=\prod_{n=-(k-1),2}^{k-1}{(y_n-Y)},\quad y_n(t_n)=2\e^qu^q\cosh{(\pi t_n/p)},\quad t_n=t+inq\ .
\end{equation}
In order to describe the boundary created by $F_k(Y)$ in the double scaling limit, we will consider the following matrix correlators,
\begin{equation}\label{def_mixed}
\mathcal{M}_{l,k}(\{x\}_l,\{y\}_k)=\la\tr\dfrac1{F_l(X)}\dfrac1{F_k(Y)}\ra.
\end{equation}
The expression of $\mathcal{M}_{l,k}$ is worked out in the continuum limit in the appendix \ref{app_C}. This correlator is seen to reproduce the Liouville boundary two point function of $B_{p-k,l}$ operators with boundary parameters $s$ and $t-ipq$, which is one of the main results of this paper.

We first consider the case $l=1$ where the left boundary has matter BC $(1,1)$. The matter bcc operator inserted is $\psi_{p-k,1}$ so that the right boundary, created by $F_k(Y)$, must have $(p-k,1)$ matter BC. This result is consistent with the study of ratios of $B_{1,1}$ one point functions (see section \refOld{sec_sym_FZZT}). We notice that this set of boundary conditions is also closed under fusion.

We are now able to interpret the double scaling limit of the correlators $\mathcal{M}_{l,k}$ with any $l$ as a disc with two boundaries with boundary parameters $s$ and $t-ipq$ related respectively to the cosmological constants $\xi$ (usual) and $\z$ (dual). The matter wears boundary conditions $(1,l)$ on the left side and $(p-k,1)$ on the right side. Between both, the bcc operator $B_{p-k,l}$ is inserted. This operator is the only one that can be inserted between $(1,l)$ and $(p-k,1)$ matter BC. Varying $k$ and $l$, we can realize all the operator dimensions belonging to the Kac table.

It is also possible to consider the matrix correlators made of $F_l(Y)$ and $F_k(Y)$,
\begin{equation}
\tilde{\Op}_{k,l}(\{x\}_l,\{y\}_k)=\la\tr\dfrac{1}{F_l(Y)}\dfrac{1}{F_k(Y)}\ra
\end{equation}
The description of this quantity can be derived from the symmetry that exchanges $p$ and $q$, reversing $X$ and $Y$ matrices. The correlator $\tilde{O}_{k,l}$ thus represents a disc with $(p-l,1)$ and $(p-k,1)$ matter BC and dual boundary cosmological constants. Between the two boundaries, the operator $B_{2p-k-l-1,1}$ is inserted.\\

A priori, the relation \ref{rel_FZZT_2MM} allows to construct boundaries with any matter BC $(k,l)$. Unfortunately, a naive application of our method does not work, due to the independence of the cosmological constants $\xi(s_{mn})$ in $n$ (or similarly of $\z(s_{mn})$ in $m$). This is not a problem in Liouville gravity because the Liouville BC is really characterized by the boundary parameter $s_{mn}$, instead of the cosmological constant $\xi(s_{mn})$, due to the quantum corrections \cite{Teschner2000}. On the contrary, the 2MM correlators with $(1,1)$ matter BC depends on $s_{mn}$ only through $\xi(s_{mn})$.

To be more precise, the problem is well defined at the level of the spectral curve. There, the equation $E(\xi,\o)=0$ gives $q$ solutions in $\xi$ at fixed $\o$, and $p$ solutions in $\o$ at fixed $\xi$:
\begin{align}
\begin{split}
&\xi_m=2\cosh\left[\dfrac{\pi}{q}(s+imp)\right],\quad m=-(q-1)\cdots (q-1),2\\
&\o_n=2\cosh\left[\dfrac{\pi}{p}(s+inq)\right],\quad n=-(p-1)\cdots (p-1),2.
\end{split}
\end{align}
We thus have determined $q$ values of $\xi$ and $p$ values of $\o$ such that $E(\xi_m,\o_n)=0$. It is then possible to choose $l$ contiguous solutions for $\xi$, i.e. living on adjacent sheets, and $k$ solutions for $\o$, centered on $\xi_0$, $\o_0$, in order to construct a boundary with $(k,l)$ matter BC. For instance, the boundary one point function will be given by
\begin{equation}\label{1pt}
\p_\xi Z_{k,l}=\sum_{m=-(l-1),2}^{l-1}{\sum_{n=-(k-1),2}^{k-1}{\o_n\dfrac{d\xi_m}{d\xi}}}=(-1)^{l+k}[l]_+[k]_-\o(\xi)\ ,\quad [k]_\pm=\dfrac{\sin{(\pi\bL^{\pm2}k)}}{\sin{(\pi\bL^{\pm2})}}\ ,
\end{equation}
with $\xi=2u^p\cosh{(\pi s/q)}$ and $\o(\xi)=2u^q\cosh(\pi s/p)$. We verify that the quotient $\p_\xi Z_{k,l}/\o(\xi)$ corresponds to the matter boundary one point function contributions, expressed as a ratio of the transformation matrix of characters \cite{Runkel1999},
\begin{equation}
\dfrac{S_{(k,l)(1,1)}}{S_{(1,1)(1,1)}}=(-1)^{l+k}[l]_+[k]_-\ .
\end{equation}
In a similar construction, the disc with two boundaries, cosmological constants $\xi$ and $\z$, and matter BC respectively $(k,l)$ and $(1,1)$ is obtained from the following expression,
\begin{equation}
\mathcal{D}_{(k,l),(1,1)}(\xi,\z)=\dfrac{\prod_{n=-(k-1),2}^{k-1}{\left[\o(\z)-(-1)^{l-1}\o_n\right]}}{\prod_{m=-(l-1),2}^{l-1}{\left[\z-(-1)^{k-1}\xi_m\right]}}\ .
\end{equation}
It is easy to check that it indeed gives the Liouville boundary two point functions of $B_{k,l}$ operators. However, in the 2MM settings, only the ``physical solution'' $\o_0$ among the $p$ possibilities for the resolvent is selected, due to the asymptotic requirement before the continuum limit, $W(x)\sim \k^{-2}/x$. Nevertheless, $Z_{kl}$ and $\mathcal{D}_{(k,l),(1,1)}$ might be obtained as a proper matrix model correlator.


\subsection{Symmetries among FZZT branes}\label{sec_sym_FZZT}
In this section, we investigate the realization of various symmetries of the FZZT branes within the matrix model. The following discussion can be applied to both one and two matrix models.

Let us first concentrate on the equivalence between matter boundary conditions $(k,l)$ and $(p-k,q-l)$. In the appendix, we show that the correlation function $\mathcal{M}_{k,l}$ can be identified either to the Liouville two point function with a charge $\b_{p-k,l}$, or to the inverse of the Liouville two point function with a charge $\b_{k,q-l}$. To understand this fact, we first notice that if the bcc operators have the same dimension, $h_{k,l}=h_{p-k,q-l}$, the Liouville momentum is reversed under the replacement $(k,l)\to(p-k,q-l)$: $P_{k,l}=-P_{p-k,q-l}$. This sign is normally absorbed by the absolute value in \ref{L_charge}. But if we take the alternative dressing for $B_{k,q-l}$, we obtain the relation $\b_{p-k,l}^++\b_{k,q-l}^-=Q$. This explains the inversion of the Liouville boundary two point function which is known to satisfy the reflection relation $d(\b|s,t)d(Q-\b|s,t)=1$ \cite{Fateev2000}.

In \cite{Hosomichi2008} was also suggested another symmetry relating the FZZT branes,\footnote{Note that the Cardy states $|k,l>$ and $|p-k,l>$ have the same boundary entropy up to a sign: $g(|k,l>)=(-1)^{p+q-1}g(|p-k,l>)$.}
\begin{equation}\label{sym_FZZT}
|s;k,l>=-|s-ipq;p-k,l>\ .
\end{equation}
Using the decomposition \ref{rel_FZZT_2MM}, the previous statement is equivalent to
\begin{equation}\label{sym_FZZT2}
\sum_{m=-(q-1),2}^{q-1}{|s+imp;1,1>}=\sum_{n=-(p-1),2}^{p-1}{|s+inq;1,1>}=0\ .
\end{equation}
These linear combinations are exactly those arising in the formal decomposition of $|s;1,q>$ and $|s;p,1>$. In the matrix model, we see from \ref{1pt} that the boundary one point functions $\p_\x Z_{kl}$ vanish whenever $k=q$ or $l=p$, due to the property $[q]_+=[p]_-=0$, which supports \ref{sym_FZZT2} and the existence of the symmetry \ref{sym_FZZT}. As a consequence of \ref{sym_FZZT}, the matrix operator $F_k(Y)$ can be equivalently seen as creating a boundary with $(k,1)$ matter BC and Liouville boundary parameter $t$, or $(p-k,1)$ and $t-ipq$.

This dual intepretation of the matrix operator $F_k(Y)$ is also seen at the level of one point functions. Let us consider the quotient of the matrix correlators $\tOp_k(\{y\}_k)=\frac{g}{N}\p_y\la\tr\log{F_k(Y)}\ra$ and $\o(x)$. Neglecting the non universal terms, we can write
\begin{equation}\label{rat_1pt}
\dfrac{\tOp_k(\{y\}_k)}{\o(x)}=(-1)^{k-1}[k]_-\dfrac{\tw(y)}{\o(x)}\ .
\end{equation}
The RHS contains a product of matter and Liouville contributions. The matter part is simply the ratio of matter $\psi_{1,1}$ one point functions with $(k,1)$ and $(1,1)$ boundary conditions,
\begin{equation}
\dfrac{\la\psi_{1,1}\ra_{(k,1)}}{\la\psi_{1,1}\ra_{(1,1)}}=\dfrac{S_{(k,1)(1,1)}}{S_{(1,1)(1,1)}}=(-1)^{k-1}[k]_-\ .
\end{equation}
The second contribution in the RHS of \ref{rat_1pt}, the ratio of the resolvents $\tw(\z)/\o(\xi)$ is the quotient of the Liouville boundary one point functions with a usual and a dual boundary cosmological constants: $t=ps/q=\bL^2 s$. Thus, $\tOp_k(\{y\}_k)$ gives in the continuum limit a disc with insertion of a $B_{1,1}$ boundary operator, with $(k,1)$ matter BC and $t$ boundary parameter. But the quotient \ref{rat_1pt} allows another interpretation when rewritten as
\begin{equation}
\dfrac{\tOp_k(\{y\}_k)}{\o(x)}=(-1)^{p-k-1}[p-k]_-(-1)^{p+q-1}\dfrac{\tw(y)}{\o(x)}=-(-1)^{p-k-1}[p-k]_-\dfrac{\tw(\tilde{y})}{\o(\tilde{x})},
\end{equation}
where we used the property $[p-k]_-=(-1)^{q-1}[k]_-$ and introduced the cosmological constants $(\tilde{x},\tilde{y})$ with a shifted boundary parameter: $\tilde{x}=\e^p\xi(s-ipq)$, $\tilde{y}=\e^q\z(t-ipq)$. In this setting, $\tOp_k(\{y\}_k)$ is seen as a disc boundary one point function with $(p-k,1)$ matter BC and $t-ipq$ boundary parameter, in agreement with the symmetry \ref{sym_FZZT}. Taking $k=1$, we recover the previous interpretation of the resolvant $\tw(y)$.

\section{Conclusion}
We briefly summarize the main achievements of this paper and provide some comments about the further considerations. In the first section, we turned on an new light on the construction of the matrix boundaries in the 1MM. Starting from a relation among FZZT branes, we introduced the matrix operator $F_l$. We showed that suitably inserted in a disc correlator, $F_l$ creates a boundary with matter BC $(1,l)$. Next, we considered the matrix correlator $\Op_{l,k}$ with boundaries $F_l$ and $F_k$ and gave an explicit proof of the recursion relation relating the critical parts of $\Op_{l+1,k}$ and $\Op_{k,l+1}$ \ref{recursion}. This relation provided the identification of $\Op_{l,k}$ to a disc with $(1,l)$ and $(1,k)$ matter boundary conditions, and $B_{1,l+k-1}$ bcc operators inserted. It is noted that the construction of all the matrix correlators describing the insertion of a different bcc operator between these two matter BC is one of the main remaining open questions. We believe that our analysis can easily be extended to more general correlators using the linearization provided by the relation \ref{rel_sum_prod}.

The previous approach was subsequently applied to the 2MM and the boundaries with $(1,l)$ and $(p-k,1)$ matter boundary conditions were constructed, using respectively $F_l(X)$ and $F_{k}(Y)$ matrix operators. As a crosscheck, the matrix correlator with $F_l(X)$ and $F_k(Y)$ boundaries was identified to the disc with $(1,l)$ and $(p-k,1)$ matter BC and $B_{p-k,l}$ bcc operators, in agreement with the CFT predictions. We also clarified the realization of the symmetry $(k,l)\to(p-k,q-l)$ within the matrix model, explaining how it is related to the reflection property of the boundary Liouville two point function. We eventually verified that the symmetry among FZZT branes with matter BC $(k,l)$ and $(p-k,l)$ discovered recently in \cite{Hosomichi2008} is satisfied by the matrix boundaries. This symmetry allows to see the $F_k(Y)$ boundary as wearing $(k,1)$ matter boundary conditions and $t$ Liouville boundary parameter instead of $(p-k,1)$ and $t-ipq$. The generalization to any $(k,l)$ boundary was also discussed but the corresponding explicit matrix correlator needs further research.

As mentioned in the introduction, to study the boundary RG flows, we have to perturb away from the critical points in the space of couplings. These perturbations correspond to the introduction of boundary operators that do not change the BC. Those remain to be constructed. Moreover, going away from the conformal point is a difficult task: The identification of the perturbing coupling of the matrix model and the Liouville gravity is far from being obvious. The main complexity lies in the appearance of contact terms reflecting the ambiguity of the regularized product of operators at the same points \cite{Moore1991}. These terms lead to a non trivial transformation between the bulk matrix and gravity couplings, known as the Belavin-Zamolodchikov transform \cite{Belavin2009}. If the transformation has been worked out at the level of the bulk one-point function on the disk \cite{Belavin2010}, it still remains to be derived for the boundary couplings.

Let us also mention that the boundaries constructed here for the one and two matrix models can also be applied to other models, such as the $O(n)$ or ADE models. It would be very interesting to find out the interpretation of this construction in terms of the corresponding statistical models on the fluctuating lattice. The symmetries observed above could also be investigated.

\acknowledgments
The authors would like to thank I. Kostov for valuable discussions, and also the organizer of the workshop ``Finite-Size Technology in Low-Dimensional Quantum Systems'' held in Benasque where part of this work has been completed. The work is partially supported by the National Research Foundation of Korea (KRNF) grant funded by the Korea government (MEST) 2005-0049409 and R01-2008-000-21026-0 (R).

\appendix

\section{One matrix model}
\subsection{Technical preliminaries}
In the appendices, we will make an extensive use of the formula
\begin{equation}\label{rel_sum_prod}
\prod_{j=1}^l{\dfrac{1}{x_j-z}}=\sum_{j=1}^l{\dfrac1{x_j-z}\prod_{i\neq j}{(x_i-x_j)^{-1}}}
\end{equation}
which can be easily proven by noting that the RHS is just the decomposition of the LHS as a sum over its poles $z=x_j$ with the proper residue. This formula can be applied safely to a matrix as every factor of the product commutes. Note that the formula holds for any set of variables $x_j, j=1\cdots l$.

The equation \ref{rel_sum_prod} can be further exploited by considering its asymptotic when $z\to\infty$. Indeed, expanding the RHS and the LHS separately at infinity, and equating both series, one can show that the expressions
\begin{equation}\label{prop_pn}
p_a(x)=\sum_{j=1}^l{x_j^a\prod_{i\neq j}{(x_i-x_j)^{-1}}}\ ,\qquad a\in\mathbb{Z}\ ,
\end{equation}
are vanishing for $0\leq a<l-1$ and are polynomial when $a\geq l-1$. In particular, $p_{l-1}(x)=(-1)^{l-1}$.


\subsection{One boundary matrix correlators}\label{app_A}
As an example of application for the previous relations, we investigate the matrix correlators
\begin{equation}
\Op_l^{(k)}(\{x\}_l)=\dfrac{g}{N}\la\tr{\dfrac{1}{F_l(M)}M^{k-1}}\ra\ ,\qquad 1\leq k\leq l\ .
\end{equation}
The relation \ref{rel_sum_prod} allows to expand the ratio $1/F_l$ into
\begin{equation}
\Op_l^{(k)}(\{x\}_l)=\sum_{n=-(l-1),2}^{l-1}{\prod_{j\neq n}{(x_j-x_n)^{-1}}\ \dfrac{g}{N}\la\tr\dfrac1{x_n-M}M^{k-1}\ra}.
\end{equation}
Then, we recursively make use of the identity
\begin{equation}
\dfrac{1}{x_n-M}M^{k-1}=\dfrac{x_n}{x_n-M}M^{k-2}-M^{k-2}
\end{equation}
in order to decrease the power of the matrix $M$ inside the correlator. The second term of the RHS always give a zero contribution because of the vanishing of $p_a$ when $a<l-1$. We end up with
\begin{equation}
\Op_l^{(k)}(\{x\}_l)=\sum_{n=-(l-1),2}^{l-1}{\prod_{j\neq n}{(x_j-x_n)^{-1}}x_n^{k-1}\o(x_n)}+(\text{n.u.}).
\end{equation}
For the critical choice $x_n\in\{x\}_l$, $\o(x_n)=(-1)^{l-1}\o(x)$ factorizes and the critical part of $\Op_l^{(k)}$ is vanishing, unless $k=l$: $\Op_l^{(k)}(\{x\}_l)=\d_{kl}\o(x)+(\text{n.u.})$.

\subsection{Demonstration of the recursion relation among 1MM correlators}\label{app_D}
In this appendix, we compute the following two-point function,
\begin{align}
{\cal O}_{l,k}(\{x\}_l,\{y\}_k)=(-1)^{k+l}
\dfrac{g}{N}\left\langle
{\rm tr}
\left(
\prod_{m=-(l-1),2}^{l-1}
\frac{1}{M-x_m}
\prod_{n=-(k-1),2}^{k-1}
\frac{1}{M-y_n}
\right)
\right\rangle,
\label{Okl}
\end{align}
We first rewrite $\Op_{l,k}$ as a ratio of determinants,
\begin{align}\label{Goro_1}
{\cal O}_{l,k}= (-1)^{l+k-1}
\frac{\tilde{\Delta}_{l,k}}{\Delta_{l,k}}+(\text{n.u.}),
\end{align}
where $\tilde{\Delta}_{l,k}$ is given by
\begin{align}
\tilde{\Delta}_{l,k}=
{\rm det}
\left(
\begin{array}{cccccccc}
1 & 1 & \cdots & 1 & 1 & 1 & \cdots & 1 \\
x_{l-1} & x_{l-3} & \cdots & x_{-(l-1)} & 
y_{k-1} & y_{k-3} & \cdots & y_{-(k-1)} \\
x_{l-1}^2 & x_{l-3}^2 & \cdots & x_{-(l-1)}^2 &
y_{k-1}^2 & y_{k-3}^2 & \cdots & y_{-(k-1)}^2 \\
\vdots & \vdots & & \vdots &
\vdots & \vdots & & \vdots \\
x_{l-1}^{k+l-2} & x_{l-3}^{k+l-2} & \cdots & x_{-(l-1)}^{k+l-2} &
y_{k-1}^{k+l-2} & y_{k-3}^{k+l-2} & \cdots & y_{-(k-1)}^{k+l-2} \\
\omega(x_{l-1}) & \omega(x_{l-3}) & \cdots & \omega(x_{-(l-1)}) &
\omega(y_{k-1}) & \omega(y_{k-3}) & \cdots & \omega(y_{-(k-1)}) \\
\end{array}
\right),
\label{delta tilde}
\end{align}
and $\Delta_{l,k}$ is the Vandermonde determinant for both variables $\{x_m,y_n\}$, 
\begin{align}
\Delta_{l,k}=
{\rm det}
\left(
\begin{array}{cccccccc}
1 & 1 & \cdots & 1 & 1 & 1 & \cdots & 1 \\
x_{l-1} & x_{l-3} & \cdots & x_{-(l-1)} & 
y_{k-1} & y_{k-3} & \cdots & y_{-(k-1)} \\
x_{l-1}^2 & x_{l-3}^2 & \cdots & x_{-(l-1)}^2 &
y_{k-1}^2 & y_{k-3}^2 & \cdots & y_{-(k-1)}^2 \\
\vdots & \vdots & & \vdots &
\vdots & \vdots & & \vdots \\
x_{l-1}^{k+l-1} & x_{l-3}^{k+l-1} & \cdots & x_{-(l-1)}^{k+l-1} &
y_{k-1}^{k+l-1} & y_{k-3}^{k+l-1} & \cdots & y_{-(k-1)}^{k+l-1} \\
\end{array}
\right).
\end{align}
In writing \ref{Goro_1}, we assumed that we can replace the resolvents by their critical values, up to a non universal polynomial. To show this property, we decompose both $1/F_l$ and $1/F_k$ using the formula \ref{rel_sum_prod}. The remaining matrix correlator can be written using the resolvent $W$,
\begin{align}
\begin{split}
\dfrac{g}{N}\la\tr\dfrac{1}{x_m-M}\dfrac{1}{y_n-M}\ra&=-\dfrac{W(y_m)-W(x_n)}{y_m-x_n}\\
&=-\dfrac{\o(y_m)-\o(x_n)}{y_m-x_n}-\dfrac{V'(y_m)-V'(x_n)}{2(y_m-x_n)}\ .
\end{split}
\end{align}
In order to replace $W$ by its singular part $\o$, we need to show that the following quantities are polynomials,
\begin{equation}
\mathcal{V}_{l,k}(\{x\}_l,\{y\}_k)=\sum_{m=(-l-1),2}^{l-1}{\sum_{n=-(k-1),2}^{k-1}{\dfrac{V'(y_m)-V'(x_n)}{y_m-x_n}\prod_{j\neq m}{(x_j-x_m)^{-1}}\prod_{j\neq n}{(y_j-y_n)^{-1}}}}
\end{equation}
We notice that the ratio involving the derivative of the potential is actually a polynomial of $x_m$ and $y_n$. It is thus sufficient to show that the quantities
\begin{equation}
\mathcal{C}_{ab}(\{x\}_l,\{y\}_k)=\sum_{m=(-l-1),2}^{l-1}{\sum_{n=-(k-1),2}^{k-1}{x_m^ay_n^b\prod_{j\neq m}{(x_j-x_m)^{-1}}\prod_{j\neq n}{(y_j-y_n)^{-1}}}}
\end{equation}
are polynomial. In this expression, the two sums can be factorized and we obtain the product $p_a(\{x\}_l)p_b(\{y\}_k)$, where the polynomials $p_a$ and $p_b$ have been defined in \ref{prop_pn}.\\

We now come back to the expression \ref{Goro_1} and simplify the determinants using the explicit expression of $x_m$ and $y_n$, $x_n=\e u\cosh{(\pi\bL s_n)}$ and $y_m=\e u\cosh{(\pi\bL t_m)}$ with the shifted boundary parameters $s_n=s+i\bL n$, $t_m=t+i\bL m$. We note that
\begin{align}\label{c pol}
x_n^k= (\e u)^k\cosh^k(\pi\bL s_n) = (\e u)^k2^{-(k-1)}\cosh(k\pi\bL s_n)+ O(x_n^{k-1}).
\end{align}
In \ref{delta tilde}, we eliminate the terms of $O(x_j^{n-1})$ in \ref{c pol} by adding other rows in the determinant multiplied by some constants. Then, if we subtract the columns of $x_m$ and $y_n$ from those of $x_{m+2}$ and $y_{n+2}$ respectively and expand the determinant with respect to the first and the last rows, we get
\begin{align}
&\tilde{\Delta}_{l,k}=c_0(\e u)^{(k+l-2)(k+l-1)/2}\left[(-1)^{k-1}\omega(y)-(-1)^{l-1}\omega(x)\right]
\prod_{j=1}^{k+l-2} 2\sin (\pi b^2 j)
\nonumber\\
& \times {\rm det}
\left(
\begin{array}{cccccc}
\sinh(\pi\bL s_{l-2}) & \cdots & \sinh(\pi\bL s_{-(l-2)}) & 
\sinh(\pi\bL t_{k-2}) & \cdots & \sinh(\pi\bL t_{-(k-2)}) \\
\sinh(2\pi\bL s_{l-2}) & \cdots & \sinh(2\pi\bL s_{-(l-2)}) &
\sinh(2\pi\bL t_{k-2}) & \cdots & \sinh(2\pi\bL t_{-(k-2)}) \\
\vdots &  & \vdots &
\vdots &  & \vdots \\
\sinh(a\pi\bL s_{l-2}) & \cdots & \sinh(a\pi\bL s_{-(l-2)})&
\sinh(a\pi\bL t_{k-2}) & \cdots & \sinh(a\pi\bL t_{-(k-2)})\\
\end{array}
\right),
\label{Okl2}
\end{align}
with $c_0=(-1)^k2^{-\frac{(k+l-2)(k+l-3)}{2}}$ and $a=k+l-2$. At this point, we need the relation $\sin n \theta = U_{n-1}(\cos \theta) \sin \theta $ where $U_{n}(x)$ is the 
Chebyshev polynomial of the second kind, given by $U_{n}(x)=(2x)^{n}+{\cal O}(x^{n-1})$. Applying this relation to the determinant in \ref{Okl2}, one can again keep only the highest order term in $U_{n}(x)$. Then, we find that $\tilde{\Delta}_{l,k}$ is given by
\begin{align}
\tilde{\Delta}_{l,k}=&(-1)^l(\e u)^{k+l-2}\left[(-1)^{k-1}\omega(y)-(-1)^{l-1}\omega(x)\right]
\Delta_{l-1\ k-1}\\
\times
&\prod_{j=1}^{k+l-2} 2\sin (\pi \bL^2 j)
\prod_{m=-(l-2)}^{l-2} \sinh (\pi\bL s_m)
\prod_{n=-(k-2)}^{k-2} \sinh (\pi\bL t_n),
\end{align}
Using this formula, one can find that the ratio $\frac{{\cal O}_{l,k+2}}{{\cal O}_{l+2\ k}}$ is given by a constant independent of $s$ and $t$,
\begin{align}
\frac{{\cal O}_{l,k+2}}{{\cal O}_{l+2,k}}
=\frac{\sinh(\pi\bL t_k) \sinh(\pi\bL t_{-k})}
{\sinh(\pi\bL s_{l}) \sinh(\pi\bL s_{-l})}
\frac{\Delta_{l-1,k+1}\Delta_{l+2,k}}
{\Delta_{l+1,k-1} \Delta_{l,k+2}}
=
\frac{[l+1]_+[l]_+}{[k+1]_+[k]_+},
\label{recursion 2}
\end{align}
The relation \ref{recursion 2} relates ${\cal O}_{l,k}$ with $k$ even to ${\cal O}_{k+l-2,2}$ and that with $k$ odd to ${\cal O}_{k+l-1,1}$. However, it is easy to show that ${\cal O}_{m,2}$ is also proportional to ${\cal O}_{m+1,1}$ as ${\cal O}_{m,2}=[m]_+{\cal O}_{m+1,1}$, so that ${\cal O}_{l,k}$ with any $k, l$ is finally reduced to ${\cal O}_{k+l-1}$ in the continuum limit,
\begin{align}
{\cal O}_{l,k}=
\frac{\prod_{j=k}^{k+l-2}[j]_+}
{\prod_{j=1}^{l-1}[j]_+}
{\cal O}_{k+l-1\ }.
\end{align}
This formula implies the simple recursion relation \ref{recursion}.

\section{Two matrix model}
\subsection{Derivation of the spectral curve and the mixed trace}\label{app_B}
In order to derive the expression of the spectral curve, we use the loop equation technique which is based on the invariance of the matrix measure. The first identity follows from
\begin{equation}
\int{dXdY\ \tr\dfrac{\p}{\p Y}\left(\dfrac1{x-X} e^{-\frac{N}{g}S[X,Y]}\right)}=0\ ,
\end{equation}
where we denoted the matrix model action $S[X,Y]=\tr\left[V(X)+U(Y)-XY\right]$. The previous equation leads to a relation between the following correlators,
\begin{equation}\label{expr_res}
xW(x)+g=-\dfrac{g}{N}\la\tr\dfrac1{x-X}U'(Y)\ra,
\end{equation}
where we used $\la\tr 1\ra=N$.

We need a second relation, obtained from a slightly more complicated expression,
\begin{equation}
\int{dXdY\ \tr\dfrac{\p}{\p X}\left(\dfrac1{x-X}\dfrac{U'(y)-U'(Y)}{y-Y} e^{-\frac{N}{g}S[X,Y]}\right)}=0\ ,
\end{equation}
which in the planar limit, and using \ref{expr_res}, reduces to
\begin{equation}\label{expr_u}
\left[y-\o(x)\right]\dfrac{g}{N}\la\tr\dfrac1{x-X}\dfrac{U'(y)-U'(Y)}{y-Y}\ra=(x-U'(y))W(x)-P(x,y)+g\ ,
\end{equation}
introducing the polynomial
\begin{equation}
P(x,y)=\dfrac{g}{N}\la\tr\dfrac{V'(x)-V'(X)}{x-X}\dfrac{U'(y)-U'(Y)}{y-Y}\ra.
\end{equation}
Taking the equation \ref{expr_u} at the point $y=\o(x)$, the LHS of \ref{expr_u} disappear and we end up with
\begin{equation}\label{mt_pol}
E(x,y)=(x-U'(y))(y-V'(x))-P(x,y)+g=0
\end{equation}
The polynomial $E(x,y)$ is of degree $\text{deg }V$ in $x$ and $\deg U$ in $y$, it is called the spectral curve. As demonstrated above, it vanishes when $y=\o(x)$, or by symmetry $x=\tw(y)$.\\

In order to derive the expression of the spectral curve in the double scaling limit, we will suppose the simplest choice for the potentials $V$ and $U$ leading to the $(p,q)$ critical point: The potential $V$ and $U$ are respectively of degree $q$ and $p$. As consequence, $E(x,y)$ is of degree $q$ in $x$ and $p$ in $y$, and the equation $E(x,y)=0$ leads to $q$ solutions $x_m$ at fixed $y$. Thus, $E(x,y)$ can be written as
\begin{equation}
E(x,y)=P(y)\prod_{m=0}^{q-1}{(x-x_m)}\ ,
\end{equation}
where $P(y)$ is a polynomial in $y$ which will be determined below. Using the parameterization \ref{param_xy} in the continuum limit, it is easy to see that the condition $\o(x_m)=y$ indeed provide $q$ solutions,
\begin{equation}
x_m=2(\e u)^p\cosh(\pi s_m/q),\qquad s_m=t+2imp\ ,
\end{equation}
where $y(t)=2(\e u)^q\cosh(\pi t/p)$ and $m$ runs from zero to $q-1$. This expression allows us to compute explicitely the product
\begin{align}
\begin{split}
\prod_{m=0}^{q-1}{(x-x_m)}&=2^q(\e u)^{pq}\prod_{m=0}^{q-1}{\left[\cosh\left(\dfrac{\pi s}{q}\right)-\cosh\left(\dfrac{\pi}{q}(t+2imp)\right)\right]}\\
&=2(\e u)^{pq}\left[\cosh{(\pi s)}-\cosh{(\pi t)}\right]
\end{split}
\end{align}
where we used the trigonometric formula (2.15) of \cite{Hosomichi2008}. This expression can be written in terms of Chebyshev polynomials of the first kind, $T_p(\cos\th)=\cos(p\th)$. We thus obtain the following expression for the spectral curve,
\begin{equation}\label{mt_ct}
E(x,y)=2(\e u)^{pq}P(y)\left[T_q\left(\dfrac{x}{2(\e u)^p}\right)-T_p\left(\dfrac{y}{2(\e u)^q}\right)\right]\ .
\end{equation}
Since $E$ is of degree $p$ in $y$, the polynomial $P(y)$ must be a constant. This constant can be obtained by comparing the asymptotics of the expression \ref{mt_ct} and \ref{mt_pol} at $x\to\infty$. Under a suitable rescaling of the matrices such that $V(x)\sim -\frac{1}{q}x^q+\cdots$, we can take $P(y)=1$, recovering the expression given in \ref{spec_cont}.

\subsection{Expression for the mixed trace}\label{app_MT}
To derive the expression of the mixed trace, we start from the following loop equation,
\begin{equation}
\int{dXdY\ \tr\dfrac\p{\p Y}\left(\dfrac1{x-X}\dfrac1{y-Y} e^{-\frac{N}{g}S[X,Y]}\right)}=0\ ,
\end{equation}
which leads in the planar limit to 
\begin{equation}
\left[x-\tw(y)\right]\mathcal{M}(x,y)=-\dfrac{g}{N}\la\tr\dfrac1{x-X}\dfrac{U'(y)-U'(Y)}{y-Y}\ra-\tW(y)\ .
\end{equation}
Plugging the expression \ref{expr_u} we get
\begin{equation}
\mathcal{M}(x,y)=1-\dfrac{(x-U'(y))(y-V'(x))-P(x,y)+g}{(x-\tw(y))(y-\o(x))}
\end{equation}
where we recognize the spectral curve in the numerator.

\subsection{Boundary two point function}\label{app_C}
In this appendix, we study the continuum limit of the mixed trace matrix correlator $\mathcal{M}_{l,k}(\{x\}_l,\{y\}_k)$ defined in \ref{def_mixed}. Using twice the formula \ref{rel_sum_prod}, and then the expression for the mixed trace \ref{mixed_trace}, one get the double sum
\begin{align}
\begin{split}
\mathcal{M}_{l,k}(\{x\}_l,\{y\}_k)=-\sum_{m=-(l-1),2}^{(l-1)}\sum_{n=-(k-1),2}^{k-1}&\dfrac{E(x_m,y_n)}{(x_m-\tw(y_n))(y_n-\o(x_m))}\times\\
&\times\prod_{j\neq m}(x_j-x_m)^{-1}\prod_{j\neq n}(y_j-y_n)^{-1}
\end{split}
\end{align}
where we recall the notations $x_m=x(s_m)=2\e^p u^{p}\cosh{(\pi s_m/q)}$ and $y_n=y(t_n)=2\e^q u^{q}\cosh{(\pi t_n/p)}$ for the cosmological constants with shifted boundary parameters $s_m=s+imp$ and $t_n=t+inq$. The resolvents $\o$ and $\tw$ take the same values for all the parameters $x_m$ and $y_n$ respectively: $\o(x_m)=(-1)^{l-1}\o(x)$ and $\tw(y_n)=(-1)^{k-1}\tw(y)$ with $x=x(s)$ and $y=y(t)$. The spectral curve is also independent of $m$ and $n$, and can be factorized out of the sums,
\begin{equation}
\mathcal{M}_{l,k}(\{x\}_l,\{y\}_k)=-E(x,y)\equskip\sum_{m=-(l-1),2}^{(l-1)}{\sum_{n=-(k-1),2}^{k-1}{\dfrac{\prod_{j\neq m}(x_j-x_m)^{-1}\prod_{j\neq n}(y_j-y_n)^{-1}}{\left[x_m-(-1)^{k-1}\tw(y)\right]\left[(y_n-(-1)^{l-1}\o(x)\right]}}}
\end{equation}
with
\begin{equation}
E(x,y)=\e^{pq} u^{pq}\left[(-1)^{p(l-1)}\cosh{(\pi s)}-(-1)^{q(k-1)}\cosh{(\pi t)}\right].
\end{equation}
Then we use back the relation \ref{rel_sum_prod} in order to obtain a factorized form,
\begin{align}
\begin{split}
\mathcal{M}_{l,k}(\{x\}_l,\{y\}_k)&=-E(x,y)\prod_{m=-(l-1),2}^{l-1}\left[x_m-(-1)^{k-1}\tw(y)\right]^{-1}\\
&\times\prod_{n=-(k-1),2}^{k-1}\left[y_n-(-1)^{l-1}\o(x)\right]^{-1}.
\end{split}
\end{align}
We now plug in the expression for the cosmological constants and resolvent in terms of the boundary parameters,
\begin{align}
\begin{split}
\dfrac{\mathcal{M}_{l,k}(\{x\}_l,\{y\}_k)}{E(x,y)}=-c_1&\prod_{m=-(l-1),2}^{l-1}{\left[\cosh\left(\dfrac{\pi s_m}{q}\right)-(-1)^{k-1}\cosh\left(\dfrac{\pi t}{q}\right)\right]^{-1}}\\
\times&\prod_{n=-(k-1),2}^{k-1}{\left[\cosh\left(\dfrac{\pi t_n}{p}\right)-(-1)^{l-1}\cosh\left(\dfrac{\pi s}{p}\right)\right]^{-1}}
\end{split}
\end{align}
where we denoted $c_1^{-1}=2^{l+k}(\e u)^{lp+kq}$. Transforming the difference of hyperbolic cosines into a product of hyperbolic sines and reorganizing the indices, one can write
\begin{align}
\begin{split}
\dfrac{\mathcal{M}_{l,k}(\{x\}_l,\{y\}_k)}{E(x,y)}=-c_2(-1)^{kl}\prod_\pm&\prod_{m=0}^{l-1}{\left(\sinh{\left[\dfrac{\pi s_\pm}{q}+i\pi\dfrac{p}{q}\left(m-\dfrac{l-1}{2}-\dfrac{k-1}{2}\dfrac{q}{p}\right)\right]}\right)^{-1}}\\
\times &\prod_{n=0}^{k-1}{\left(\sinh{\left[\dfrac{\pi s_\pm}{p}+i\pi\dfrac{q}{p}\left(n-\dfrac{l-1}{2}\dfrac{p}{q}-\dfrac{k-1}{2}\right)\right]}\right)^{-1}}
\end{split}
\end{align}
with $s_\pm=(s\pm t)/2$ and $c_2=2^{-l-k}c_1$. Using the formula (2.15) of \cite{Hosomichi2008} in order to write the spectral curve as a product of hyperbolic sines,
\begin{align}
\begin{split}
E(x,y)&=(-1)^{q(k-1)}\e^{pq} u^{pq}\left[\cosh{(\pi s-i\pi(l-1)p-i\pi(k-1)q)}-\cosh{(\pi t)}\right]\\
&=(-1)^{q(k-1)+p-1}2^{2(p-1)}\e^{pq} u^{pq}\prod_\pm\prod_{n=0}^{p-1}{\sinh\left[\dfrac{\pi s_\pm}{p}+i\pi\dfrac{q}{p}\left(n-\dfrac{l-1}{2}\dfrac{p}{q}-\dfrac{k-1}{2}\right)\right]}
\end{split}
\end{align}
we observe that some of the terms simplify. We end up with the following ratio for $\mathcal{M}_{l,k}(\{x\}_l,\{y\}_k)$:
\begin{align}
\begin{split}\label{okl_fin}
&(-1)^{p+q+k(q-l)}c_4\prod_\pm\dfrac{\prod_{n=k}^{p-1}{\sinh\left[\dfrac{\pi s_\pm}{p}+i\pi\dfrac{q}{p}\left(n-\dfrac{l-1}{2}\dfrac{p}{q}-\dfrac{k-1}{2}\right)\right]}}{\prod_{m=0}^{l-1}{\sinh{\left[\dfrac{\pi s_\pm}{q}+i\pi\dfrac{p}{q}\left(m-\dfrac{l-1}{2}-\dfrac{k-1}{2}\dfrac{q}{p}\right)\right]}}}\\
=&(-1)^{p+q+pq+(p-k)l}c_4\prod_\pm\dfrac{\prod_{n=0}^{p-k-1}{\sinh\left[\dfrac{\pi \ts_\pm}{p}+i\pi\dfrac{q}{p}\left(n-\dfrac{l-1}{2}\dfrac{p}{q}-\dfrac{p-k-1}{2}\right)\right]}}{\prod_{m=0}^{l-1}{\sinh{\left[\dfrac{\pi\ts_\pm}{q}+i\pi\dfrac{p}{q}\left(m-\dfrac{l-1}{2}-\dfrac{p-k-1}{2}\dfrac{q}{p}\right)\right]}}}
\end{split}
\end{align}
with the irrelevant constant $c_4=2^{2(p-l-k-1)}(\e u)^{(p-k)q-lp}$. In the RHS, we introduced a shifted boundary parameter, $\ts_\pm=(s\pm \tilde{t})/2$ with $\tilde{t}=t-ipq$. The expression of the Liouville boundary two point function $d(\b|s,t)$ derived in \cite{Fateev2000} involve a ratio of double sine functions. But when the Liouville theory is coupled to a minimal model, this expression degenerates into a ratio of hyperbolic sine which can be found in \cite{Kostov2003a} (appendix A). For a Liouville charge $\b_{\rho,\s}$ given by \ref{L_charge}, and $s$ and $t$ boundary parameters, the formula reads
\begin{equation}\label{form_Kostov}
d(\b_{\r,\s}|s,t)=cu^{|P_{r,s}|\sqrt{pq}/2}\prod_\pm\dfrac{\prod_{n=0}^{\r-1}{\sinh\left[\dfrac{\pi s_\pm}{p}+i\pi\dfrac{q}{p}\left(n-\dfrac{\s-1}{2}\dfrac{p}{q}-\dfrac{\r-1}{2}\right)\right]}}{\prod_{m=0}^{\s-1}{\sinh{\left[\dfrac{\pi s_\pm}{q}+i\pi\dfrac{p}{q}\left(m-\dfrac{\s-1}{2}-\dfrac{\r-1}{2}\dfrac{q}{p}\right)\right]}}}
\end{equation}
where $c$ is a constant independent of the cosmological constants. Comparing \ref{okl_fin} and \ref{form_Kostov}, one can deduce that in the continuum limit $\mathcal{M}_{k,l}$ is proportional to the minimal Liouville gravity two point function of $B_{p-k,l}$ boundary operators with boundary parameters $s$ and $t-ipq$ corresponding respectively to a usual $\xi(s)$ and dual $\z(t-ipq)$ boundary cosmological constants.

Let us conclude with the following remark: the decomposition of the spectral curve can also be made on the dual product of hyperbolic sine,
\begin{equation}
E(x,y)=(-1)^{q(k-1)+q-1}2^{2(q-1)}\e^{pq} u^{pq}\prod_\pm\prod_{m=0}^{q-1}{\sinh\left[\dfrac{\pi s_\pm}{q}+i\pi\dfrac{p}{q}\left(m-\dfrac{l-1}{2}-\dfrac{k-1}{2}\dfrac{q}{p}\right)\right]}.
\end{equation}
In this case, we obtain another expression for the matrix correlator $\mathcal{M}_{l,k}$,
\begin{equation}
\mathcal{M}_{l,k}(\{x\}_l,\{y\}_k)\propto\prod_\pm\dfrac{\prod_{m=0}^{q-l-1}{\sinh\left[\dfrac{\pi\ts_\pm}{q}+i\pi\dfrac{p}{q}\left(m-\dfrac{q-l-1}{2}-\dfrac{k-1}{2}\dfrac{q}{p}\right)\right]}}{\prod_{n=0}^{k-1}{\sinh{\left[\dfrac{\pi\ts_\pm}{p}+i\pi\dfrac{q}{p}\left(n-\dfrac{q-l-1}{2}\dfrac{p}{q}-\dfrac{k-1}{2}\right)\right]}}}
\end{equation}
which should now be identified with the inverse of \ref{form_Kostov} for a $B_{k,q-l}$ boundary operator with again $s$ and $t-ipq$ boundary parameters.
\bibliographystyle{JHEP}
\providecommand{\href}[2]{#2}\begingroup\raggedright\endgroup

\end{document}